\documentclass[10pt,a4j]{article}
\usepackage[dvips]{graphicx}
\usepackage{mathrsfs}
\usepackage{eucal}
\usepackage{amsmath,amsthm,amssymb}
\usepackage{wrapfig}
\usepackage[dvips]{color}
\usepackage{cite}
\begin{document}
\renewcommand{\thefootnote}{$\dagger$\arabic{footnote}}
\begin{center}
\textbf{\LARGE Theory of Gel Formation}
\end{center}
\vspace{0mm}
\begin{center}
\textbf{\Large Test of the Gel Theory by Ilavsky Experiments}
\end{center}
 \vspace*{0mm}
\begin{center}
\large{Kazumi Suematsu} \vspace*{2mm}\\
\normalsize{\setlength{\baselineskip}{12pt} 
Institute of Mathematical Science\\
Ohkadai 2-31-9, Yokkaichi, Mie 512-1216, JAPAN\\
E-Mail: suematsu@m3.cty-net.ne.jp,  Tel/Fax: +81 (0) 593 26 8052}\\[10mm]
\end{center}

\begin{center}
\textbf{\Large{Summary}}\vspace*{-2mm}\\
\end{center}

\setlength{\baselineskip}{13pt} 
The theory of gelation is tested by the recent experiments in poly(urethane) network.  The result supports strongly the physical soundness of the theory.

\begin{wrapfigure}[16]{r}{5.1cm}
\vspace*{-1mm}
\includegraphics[width=5.1cm]{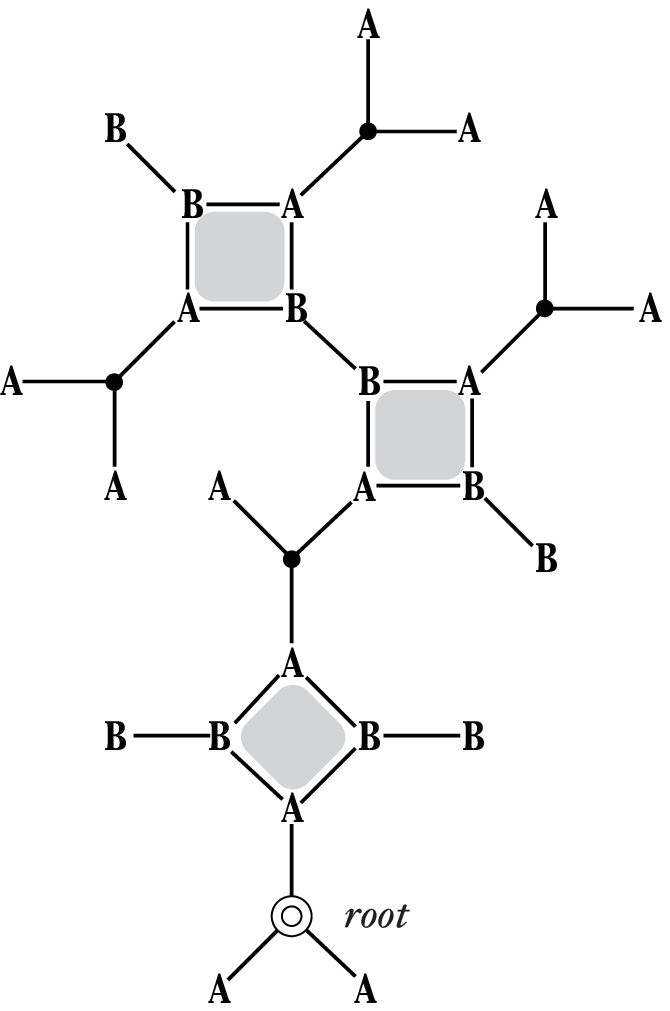} 
\\[2.7mm]\setlength{\baselineskip}{10.5pt}{\small{Fig. 1: Typical R$-$A$_{g}\hspace{0.5mm}+\hspace{0.5mm}$R$-$B$_{f-g}$ multilink polymerization for $J=4$.} }
\end{wrapfigure}
\section{Introduction}
It is shown\cite{Kazumi} that the gel point condition is
\begin{equation}
D_{A_c}=\frac{\sqrt{t^2+4s}-t}{2s}\left\{\frac{1-\mathcal{F}\displaystyle\sum\nolimits_{j}\left(1-1/j\right)\varphi_j\{\dotsm\}\gamma_f}{1-\mathcal{F}\displaystyle\sum\nolimits_{j}\varphi_j\{\dotsm\}\gamma_f}\right\},
\end{equation}
for the multilink system of the R$-$A$_g$+R$-$B$_{f-g}$ model, where 
\begin{gather}
\displaystyle\mathcal{F}=\frac{(1+\kappa)J}{4(J-1)}\left(\frac{1}{2}\ t+\sqrt{s+\left(\frac{1}{2}\ t\right)^2}\right),\\
s=(J-1)(\langle g_w\rangle-1)(\langle (f-g)_w\rangle-1)/\kappa,\\
t=\textstyle{\left(\frac{J}{2}-1\right)}\displaystyle\left\{(\langle g_w\rangle-1)+(\langle (f-g)_w\rangle-1)/\kappa\right\}.
\end{gather}
In eq. (1), $D_{A}$ denotes the extent of reaction of A  functional units (FU) $g$ and $f-g$ the A  and B FU (FU), respectively, the symbol $\langle( \hspace{2mm})_{w}\rangle$ the weight average quantity, $\varphi_{j}$ the relative cyclization frequency\cite{Kazumi} for $j$-chains, $\kappa=\sum_j(f-g)_jM_{B_j}/\sum_ig_iM_{A_i}$ the relative mole number of B FU to A FU, and $J$ the multiplicity that represents the number of FU necessary to form a junction point. A typical example is shown in Fig. 1 for $g=3, f-g=2$, and $J=4$.
\section{Special Solution of $J=2$}
For $J=2$, the bond formation between the same FU's is forbidden and $t=0$, so all odd terms of the sum in eq. (1) cancel out, which results in
\begin{equation}
D_{A_c}=\sqrt{\frac{1}{s}}\left\{\frac{1-\left(1+\kappa\right)\sqrt{s}\hspace{1mm}\displaystyle\sum\nolimits_{j}\left(1-1/2j\right)\varphi_{2j}\hspace{0.3mm}\gamma_f}{1-\left(1+\kappa\right)\sqrt{s}\hspace{1mm}\displaystyle\sum\nolimits_{j}\varphi_{2j}\hspace{0.3mm}\gamma_f}\right\},
\end{equation}
where $s=\left(\langle g_w\rangle-1\right)\left(\langle (f-g)_w\rangle-1\right)/\kappa$. The relative frequency of cyclization can be expressed by the incomplete Gamma function of the form:
\begin{equation}
\varphi_j=\big(d/2\pi^{d/2}\ell^{\hspace{0.3mm}d}N_{A}\big)\displaystyle\int_{0}^{d/2\nu_{\hspace{-0.3mm}j}}\hspace{-2mm}t^{\frac{d}{2}-1}e^{-t}dt \hspace{5mm}\text{for ideal molecules},
\end{equation}
where $d$ is dimension and $\ell$ the bond length. $\varphi_j$ is also the function of the quantity $\nu_{j}$ defined by the end-to-end distance $\langle r_{j}^{2}\rangle=\nu_{j}\hspace{0.3mm}\ell^{\hspace{0.3mm}2}$. For ideal chains with sufficient length, it follows that $\nu_{j}=C_{F}\hspace{0.2mm}\xi\hspace{0.3mm} j$ where $C_{F}$ denotes the Flory characteristic constant, $\xi$ the number of bonds within the repeating unit, and $j$ the number of the repeating unit. In the multilink R$-$A$_g$+R$-$B$_{f-g}$ model, the repeating unit has been defined as the length of an A$-$A or a B$-$B chain on the monomers. However, for the present case of $J=2$ it is more convenient to define the repeating unit as the length of an A$-$A$\cdot$B$-$B chain on the dimer. For this purpose, let us transform $\nu_{j}$ as $\nu_{j}=C_{F}\hspace{0.2mm}\xi\hspace{0.3mm}j=C_{F}\hspace{0.2mm}(2\hspace{0.2mm}\xi)\hspace{0.3mm}(j/2)$. The $2\hspace{0.2mm}\xi$ corresponds to the number of bonds within the newly defined repeating unit, A$-$A$\cdot$B$-$B, and the $(j/2)$ the number of the new repeating unit; i.e., $($A$-$A$\cdot$B$-$B$)_{j/2}$, which can be assigned to even terms of $\varphi_{j}$. Now one can write eq. (5) in the form:
\begin{equation}
D_{A_c}=\sqrt{\frac{1}{s}}\left\{\frac{1-\left(1+\kappa\right)\sqrt{s}\hspace{1mm}\displaystyle\sum\nolimits_{x}\left(1-1/2x\right)\varphi_{x}\hspace{0.3mm}\gamma_f}{1-\left(1+\kappa\right)\sqrt{s}\hspace{1mm}\displaystyle\sum\nolimits_{x}\varphi_{x}\hspace{0.3mm}\gamma_f}\right\},
\end{equation}
where
\begin{equation}
\varphi_x=\big(d/2\pi^{d/2}\ell^{\hspace{0.3mm}d}N_{A}\big)\displaystyle\int_{0}^{d/2\nu_{\hspace{-0.3mm}x}}\hspace{-2mm}t^{\frac{d}{2}-1}e^{-t}dt,
\end{equation}
and 
\begin{equation}
\nu_{x}=C_{F}\hspace{0.2mm}\xi'\hspace{0.3mm}x\hspace{5mm}(x=1, 2, 3, \cdots). 
\end{equation}
$\xi'$ represents the number of bonds within the new unit length, A$-$A$\cdot$B$-$B, and $x$ the number, so that $($A$-$A$\cdot$B$-$B$)_{x}$. 

If we replace $\gamma_{f}$ with the reciprocal, $\gamma$, of the monomer unit concentration\hspace{0.3mm}\footnote{\vspace{0mm}\hspace{1mm}$\displaystyle\gamma_{f}=\frac{\big(V\big)}{\big(\sum\nolimits_{i}g_{i}M_{A_{i}}+\sum\nolimits_{j}(f-g)_{j}M_{B_{j}}\big)}$, while $\displaystyle\gamma=\frac{\big(V\big)}{\big(\sum\nolimits_{i}M_{A_{i}}+\sum\nolimits_{j}M_{B_{j}}\big)}$.},\\[1mm]
\begin{equation}
\gamma_f\rightarrow\frac{\big\langle(f-g)_{n}\big\rangle+\big\langle g_{n}\big\rangle\hspace{0.3mm}\kappa}{\big\langle g_{n}\big\rangle\hspace{0.3mm}\big\langle(f-g)_{n}\big\rangle\big(1+\kappa\big)}\hspace{0.5mm}\gamma,
\end{equation}
\big(the symbol $\big\langle(\hspace{2mm})_{n}\big\rangle$ denotes the number average quantity defined, for instance, by $\big\langle g_{n}\big\rangle=\sum_{i}g_{i}M_{A_{i}}/\sum_{i}M_{A_{i}}$\big), eq. (7) reduces exactly to the previous one\cite{Kazumi}.

\section{Application to the Ilavsky Experiment}
To test the theory, let us take up the recent observation by Ilavsky and coworkers\cite{Ilavsky}: the polyaddition reaction of tris(4-isocyanatophenyl)thiophosphate (TI) and poly(oxypropylene)diol (PD) to yield a poly(urethane) network. They carried out the close investigation of the critical molar ratio, $\kappa_{c}$, for gelation at $D_{A}=1$ as a function of the dilution, $\gamma_{f}$. In that case, the respective FU's are chosen so that $\kappa=\frac{\text{[OH]}}{\text{[NCO]}}\ge 1$, and therefore $\langle g_{w}\rangle=3$ and $\langle(f-g)_{w}\rangle=2$.

To compare the theory with the experiment, we impose the constraint, $0\le D_{A_{c}}\le 1$, on eq. (7) to get
\begin{equation}
0\le\gamma_f\le \gamma_{f_{c}}=\frac{1-1/\sqrt{s}}{\big(1+\kappa\big)\displaystyle\sum\nolimits_{x}\left(-1+\sqrt{s}+1/2x\right)\varphi_{x}}.
\end{equation}
\noindent Note that $s=\left(\langle g_w\rangle-1\right)\left(\langle (f-g)_w\rangle-1\right)/\kappa$ is also a function of $\kappa$.

To evaluate the value of $\varphi_{x}$, let us introduce the new quantities: the standard bond length, $\ell_{s}$, and the effective bond number, $\xi_{e}$\cite{Kazumi}. Note that a real chain has the end-to-end distance of the form:
\begin{equation}
\langle r_{x}^{2}\rangle_{\varTheta}=C_{F}\hspace{0.3mm}x\sum_{i}^{\xi'}\ell_{i}^{\hspace{0.3mm}2}=C_{F}\hspace{0.3mm}\xi'\hspace{-0.1mm}x\hspace{0.3mm}\bar{\ell}^{\hspace{0.3mm}2},
\end{equation}
for a large $x$ in the $\varTheta$ state. In practice, however, eq. (12) is not very easy to use. As one can see below, it is convenient to transform this equation as
\begin{equation}
\langle r_{x}^{2}\rangle_{\varTheta}=\nu_{x}\hspace{0.3mm}\ell_{\hspace{-0.3mm}s}^{\hspace{0.3mm}2}=C_{F}\hspace{0.2mm}\xi_{e}\hspace{0.3mm}x\hspace{0.3mm}\ell_{\hspace{-0.3mm}s}^{\hspace{0.3mm}2},
\end{equation}
where $\ell_{s}$ represents the C$-$O bond length (1.36 \AA) of the \textit{urethane moiety} for the present case. Now we can write eq. (8) in the form:
\begin{equation}
\varphi_x=\big(d/2\pi^{d/2}\ell_{s}^{\hspace{0.3mm}d}N_{A}\big)\displaystyle\int_{0}^{d/2\nu_{\hspace{-0.3mm}x}}\hspace{-2mm}t^{\frac{d}{2}-1}e^{-t}dt,
\end{equation}
\noindent where
$\nu_{x}\hspace{0.3mm}=C_{F}\hspace{0.3mm}\xi_{e}\hspace{0.3mm}x\hspace{0.3mm}$
and $\xi_{e}=\xi'\hspace{-0.3mm}\left(\bar{\ell}^{\hspace{0.3mm}2}/\ell_{\hspace{-0.3mm}s}^{\hspace{0.3mm}2}\right)$. In this way, we have determined all the parameters, as shown in Table 1. Making use of these values, we have carried out the parametric plot of eq. (11).  Experimental points ({\Large$\diamond$}) were recalculated according to the equation:\\[1mm]
\newpage
\begin{table}[h]
\caption{Parameters for the TI$-$PD Branched Poly(urethane)}
\begin{center}
\begin{tabular}{c c c}\hline\\[-2mm]
molecular weight & \hspace{3mm} unit & \hspace{3mm} $\text{TI}=465, \text{PD}=400$\\[1mm]
$\langle g_{w}\rangle$ & \hspace{3mm} & \hspace{3mm} 3 \\[1mm]
$\langle(f-g)_{w}\rangle$ & \hspace{3mm} & \hspace{3mm} 2 \\[1mm]
$d$ & \hspace{3mm} & \hspace{3mm} 3 \\[1mm]
$C_{F}$ & \hspace{3mm} & \hspace{3mm} 4.3\\[1mm]
$\xi_{e}$ & \hspace{3mm} & \hspace{3mm} 62\\[1mm]
$\ell_{s}$ & \hspace{3mm} \AA & \hspace{3mm} 1.36 \\[2mm]
cyclization frequency &  &  \\[1.5mm]
$\sum\nolimits_{x=1}^{\infty}\varphi_{x}$ & \hspace{3mm} $mol/l$ & \hspace{3mm} 0.1306\\[2mm]
$\sum\nolimits_{x=1}^{\infty}\varphi_{x}/2x$ & \hspace{3mm} $mol/l$ & \hspace{3mm} 0.03366\\[2mm]
\hline\\
\end{tabular}
\end{center}
\end{table}

\begin{wrapfigure}[7]{r}{6cm}
\vspace{0mm}
\includegraphics[width=6cm]{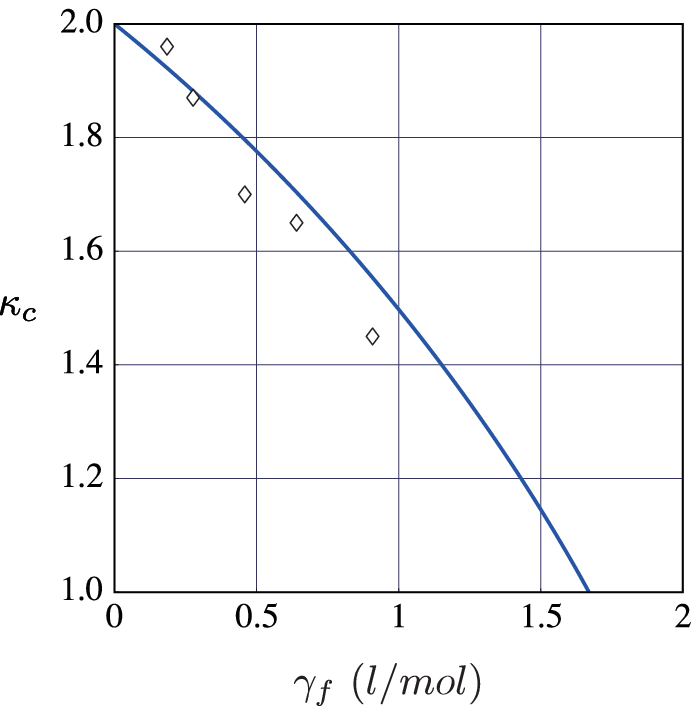}
\\[2mm]\setlength{\baselineskip}{10pt}{\small Fig. 2: $\kappa_{c}$ plot as a function of $\gamma_{f}$.\\ Solid line: theoretical line by eq. (11); {\large$\diamond$}: experimental points by Ilavsky and coworkers\cite{Ilavsky}. }
\end{wrapfigure}
\begin{equation}
\gamma_{f}=\frac{\langle(f-g)_{n}\rangle\langle m_{A,n}\rangle+\langle g_{n}\rangle\langle m_{B,n}\rangle\kappa}{1000\hspace{0.3mm}\varrho\hspace{0.3mm}v\langle g_{n}\rangle\langle(f-g)_{n}\rangle(1+\kappa)},
\end{equation}

\noindent where $\varrho$ is the density of polymer ($g/ml$), $v$ the volume fraction, and $\langle m_{A,n}\rangle$ and $\langle m_{B,n}\rangle$ are the number average molecular weights of TI and PD monomers, respctively, which are given in Table 1. In Fig. 2, the solid line is theoretical line by eq. (11) and experimental points by Ilavsky and coworkers\cite{Ilavsky}. As one can see, the observations support strongly the theory.

Recall that $C_{F}$ is not constant, but increases gradually with increasing $x$, asymptotically approaching a constant as $x\to \infty$\cite{Flory}. Thus it is important to notice that we tend to overestimate the $C_{F}$ value and thus underestimate the total amount of rings. This may explain the small difference in Fig. 2 between the theoretical line and the experimental points. Quite conversely, if the theory is exact, there is a possibility that one can gain the structural information of various branched polymers by fitting the theoretical line to the observed points.\\[1mm]


\end{document}